\documentclass[aps,prd,onecolumn,11pt,notitlepage,nofootinbib,superscriptaddress,showkeys,letterpaper]{revtex4-1}
\usepackage{amstext,amsmath,amssymb,amsfonts,bbm, amsthm}
\usepackage[latin1]{inputenc}
\usepackage{fancyhdr}
\usepackage{graphicx}
\usepackage{hyperref}
\usepackage{multirow}

\usepackage{epstopdf}
\usepackage{rotating}
\usepackage{cleveref}

\usepackage{float}
\usepackage{placeins}

\usepackage{cases}
\usepackage{tabls}

\usepackage{subfigure}

\usepackage{amsthm}

\usepackage{tocvsec2}

\usepackage{color}

\def\beq{\begin{equation}}
\def\be{\begin{equation}}
\def\ee{\end{equation}}
\def\bes{\begin{eqnarray}}
\def\ees{\end{eqnarray}}

\begin{document}

\title{\large \bf A non-equilibrium formulation of food-security resilience}

\author{Matteo Smerlak}
\affiliation{Perimeter Institute for Theoretical Physics, 31 Caroline St.~N., Waterloo ON N2L 2Y5, Canada}

\author{Bapu Vaitla}\email{To whom correspondence should be addressed. E-mail: vaitla@hsph.harvard.edu}
\affiliation{Feinstein International Center, Tufts University, 114 Curtis Street, Somerville MA 02144, USA}
\affiliation{T.H. Chan School of Public Health, Harvard University, 677 Huntington Avenue, Boston MA 02115, USA}
\date{\small\today}

\begin{abstract}
Resilience, the ability to recover from adverse events (``shocks''), is of fundamental importance to food security. This is especially true in poor countries, where basic needs are frequently threatened by economic, environmental, and health shocks. An empirically sound formalization of the concept of food security resilience, however, is lacking. Here we introduce a general framework for quantifying resilience based on a simple definition: a unit is resilient if (a) its long-term food security trend is not deteriorating and (b) the effects of shocks on this trend do not persist over time. Our approach can be applied to any food security variable for which high-frequency time-series data is available, can accommodate any unit of analysis (e.g., individuals, households, countries), and is especially useful in rapidly changing contexts wherein standard equilibrium-based economic models are ineffective. We illustrate our method with an analysis of per capita kilocalorie availability for 161 countries between 1961 and 2011. We find that resilient countries are not necessarily those that are characterized by high levels or less volatile fluctuations of kilocalorie intake. Accordingly, food security policies and programs will need to be tailored not only to welfare levels at any one time, but also to long-run welfare dynamics.
\end{abstract} 

\maketitle

\newpage
\section{Introduction}

Ahost of shocks---political conflicts, economic recessions, natural disasters, and epidemic diseases---continually threatens food security, especially in the developing world. Resilience, the ability to recover quickly from shocks, is thus of major interest to social scientific researchers and policymakers worldwide. 

The formalization of resilience, however, is not straightforward. Studies in economics, ecology, engineering, and psychology offer a diversity of approaches \cite{Barrett:2014fr, Fletcher:2013jb, Folke:2004bq}, and recent work has proposed useful conceptualizations of resilience as applied to international development  \cite{Cisse:2016vn}. Most of these methods rest on the assumption that the variables of interest---e.g., dietary intake or income---are attracted to certain values that are inherently stable, i.e., that equilibrium states exist. Resilience is consequently defined as the ability to absorb shocks without being pushed out of a desirable equilibrium state into an alternative, \emph{un}desirable equilibrium state.

The problem with this approach is twofold. First, economies and social structures, especially in poor rural areas, experience rapid, unpredictable change, and it is not clear that equilibrium-based models appropriately capture this process. Second,  even if ``real'' food security equilibrium states exist, they are very difficult to identify. 

In this paper, we introduce a statistical definition of food security resilience that, when high-frequency time series data are available, provides a non-equilibrium alternative. Our approach is based on the analysis of autocorrelation: the strength of association between past, present, and future states of a dynamical system. Strong autocorrelation is termed ``persistence,'' a concept commonly used in fields as diverse as physics \cite{Ruzmaikin:1994gs}, climatology \cite{KoscielnyBunde:1998bi}, physiology \cite{Leistedt:2007ga}, and molecular biology \cite{Peng:1994be}. A great deal of econometric work has focused on the persistence of macroeconomic time series, e.g., GNP, commodity prices, and inflation rates \cite{Nelson:1982bq}. 

With respect to food security, the importance of quantifying the persistence of the effects of shocks is clear: actors can be dragged down and kept low by past adverse events. The opposite quality could be called ``anti-persistence'': the ability to quickly disassociate from these negative impacts. For anti-persistent actors, the consequences of disaster---decreased food consumption, asset loss, psychological stress, and so on---do not endure. 

Resilience, however, is not only about independence from the past. We argue that the term should only apply to actors whose long-term welfare trend is either improving or neutral; return to a deteriorating long-term food security trend, however rapid, does not fit well with an intuitive notion of resilience. In the approach we present here, resilience is thus defined with reference to both \emph{persistence} of the effects of shocks, as measured by a chosen food security variable, and the long-term \emph{trend} of that variable.

\begin{table*}[t!]
\begin{center}
	\begin{tabular} {| c || c | c| } \hline
\textbf{Category} & \textbf{Term} & \textbf{Definition} \\ 
\hline
\multirow{4} {*} {\parbox{3cm} {Fundamental properties}} & Level $k_t$ & Value of a food security variable \\ 
& Trend $g$ & Slope of a variable; the mean increment $\mathbb{E}(\Delta_t)$ over the time series \\ 
& Volatility $\sigma$ & Quantification of mean fluctuation size in the time series  \\
& Persistence $\pi$ & Association between present and past increments in the time series \\ 
\hline
\multirow{2} {*} {\parbox{3cm}{Key derived metrics}} & Resilience & Anti-persistence of shock effects ($\pi<0$)given $g\geq 0$ \\ 
& Resistance & Lack of volatility ($\sigma<2\vert g\vert$) given $g\geq 0$\\
\hline
\end{tabular}
\end{center}
\caption{Definitions of fundamental statistical properties and key derived metrics}
\label{tabdef}
\end{table*}

More broadly, we suggest that persistence and trend are two of the four fundamental statistical properties relevant to the analysis of food security time-series data. The other two properties, \emph{level} and \emph{volatility}, are commonly analyzed in international development research. Levels of a given food security variable are evaluated with reference to a benchmark value (e.g., the level of caloric intake relative to need) or relative to the levels of comparable households, countries, or other actors. Volatility can be meaningfully interpreted by evaluating variance in the food security series relative to the slope of the long-term trend. We suggest that low volatility, also in the context of a non-deteriorating long-term trend, defines an actor as ``resistant'' to shocks. Resilience and resistance are thus distinct but complementary concepts; the former is derived from trend and persistence, the latter from trend and volatility (Table \ref{tabdef}).

\begin{figure}[t!]
\centering
\includegraphics[width=.7\linewidth]{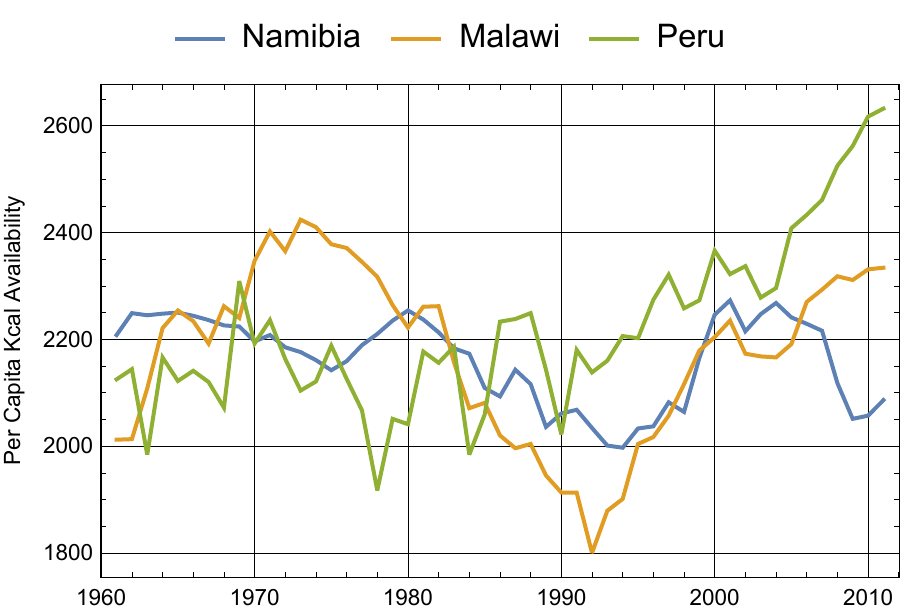}
\caption{Resilience and resistance with respect to per capita kilocalorie availability in Namibia, Malawi, and Peru between 1961 and 2011. Namibia exhibits a slightly declining long-term trend and high persistence of the effects of shocks; neither of the criteria for resilience is met. In contrast, the trend is neutral to positive for the Malawi and Peru cases. Malawi exhibits strong persistence throughout the time series. Slopes of time increments tend to be autocorrelated---rising for a decade, falling for two decades, and then rising again---and so the country is not considered resilient. The Peru series, unlike the other two countries, is very anti-persistent. The anti-persistence is seen in the relatively rapid reversals of slope; the effects of shocks do not last. Because the Peru series is both upward trending and anti-persistent, we label it resilient. All three countries, however, are volatile, and so not resistant.}
\label{fig:cdemo}
\end{figure}

Fig. \ref{fig:cdemo} illustrates these statistical features using per capita kilocalorie availability data from Namibia, Malawi, and Peru over the years 1961-2011. Note that the various trajectories in Fig. \ref{fig:cdemo} have different implications for policies and programs. For example, livelihood support interventions, such as disaster insurance and irrigation schemes, may be needed to build resilience and promote rapid recovery in Namibia and Malawi. Meanwhile, safety net interventions to strengthen resistance, such as subsidized food sales or food aid, might be appropriate to prevent food insecurity in Peru. 

\section{Non-equilibrium resilience}

As noted earlier, the chief advantage of our conceptualization of resilience, compared to other approaches, is that it does not necessitate the identification of ``equilibrium states'' or ``basins of attraction'' of food security variables \cite{Barrett:2014fr}. Dynamic economic models often focus on the relationship of current welfare to a theoretical long-run equilibrium state. In ecology, a ``basin of attraction'' is a state space within which systems are thought to maintain their fundamental identity, and ecological resilience is often defined as the ability of a system to stay within such a space \cite{Holling:1973wh}. Recent works have transferred this notion to microeconomic settings, with resilience defined as the capacity to remain, in the face of shocks, in a non-poor basin of attraction \cite{Barrett:2014fr}. Ciss\'e and Barrett \cite{Cisse:2016vn}, for example, measure resilience by evaluating the probability of attaining a well-being threshold over a given time horizon. 

The challenge with equilibrium-based analysis is to accurately make inferences from observable data about the structure of welfare dynamics. Barrett and Carter \cite{Barrett:2013dt} list some of the most important issues complicating such inferences: unstable equilibria can be difficult to identify in small samples, because their very instability makes observations around these points rare; in cluster sampling designs, homogeneity within clusters can mask equilibrium points because of regression towards the mean; and the structure of the underlying production function mapping capital inputs to outputs, and thus to well-being, may be shifting over time. More generally, estimating non-linear production functions (particularly cubic polynomials, the minimal specification for S-shaped poverty trap functions) in the presence of stochastic errors requires distributional assumptions which may be difficult to test empirically, even with fine-grained panel data. 

There are also conceptual difficulties with defining resilience with reference to a welfare threshold. First, such a metric is intrinsically unstable in the face of shocks: by definition, the probability to become poor increases, and thus resilience declines, after major asset losses, diseases, etc. This approach results in a characterization of resilience that includes both intrinsic properties of actors and exogenous conditions. Another issue with equilibrium approaches is the assumption that all actors follow the same (stochastic) dynamics: two actors with the same welfare level at a given time, must, on average, experience the same level of welfare in the future. To reconcile this assumption with observed differences in welfare trajectories requires the identification and measurement of a large number of control variables. This may be impossible, particularly if the structures of production functions vary across actors based on unobserved attributes (e.g., risk aversion). 


In summary, movement away from an equilibrium state depends on knowing the values that define that equilibrium. If one cannot know whether the pre-shock and post-shock values of a welfare variable are in equilibrium or are unstable, one cannot judge the degree to which the subsequent trajectory should be interpreted as resilient. The approach we advance in this paper does not solve this problem, but rather avoids it: we link resilience not to a target food security level or target speed of recovery, but rather evaluate it as the ability of an actor to disassociate from adverse past events while maintaining an improving or neutral long-term food security trend. This method requires high-frequency time series data; when such data is not available, alternative methods such as those developed by Ciss\'e and Barrett \cite{Cisse:2016vn} are preferable.


\section{The structure of welfare trajectories}

We now formalize the above approach. Let $k_t$ be a food security variable, measured at sufficiently high frequency across a set of actors. As a rule, we cannot assume that $k_t$ is a stationary time series: evolving policy, technology, etc., as well as endogenous effects, are likely to shift $k_t$ over time. To account for these non-equilibrium dynamics, we consider not $k_t$ directly, but rather its increment $\Delta_t=k_t-k_{t-1}$, as our dependent variable. Our approach to the analysis of food security trajectories is then based on two assumptions: $(i)$ that the increments $\Delta_t$ are (weakly) stationary, and $(ii)$ that they can be modeled by an autoregressive moving average (ARMA) $(p,q)$ process of the form 
\begin{equation}
\label{ARMA}
	\Delta_t=g+\sum_{i=1}^p\beta_i(\Delta_{t-i}-g)+\sum_{j=1}^q\theta_j\epsilon_{t-j}+\epsilon_t,
\end{equation}
where $\epsilon_t$ are independent, identically distributed shocks with standard deviation $\sigma$. \eqref{ARMA} models an actor's food security change at time $t$ as a linear function of its past changes and of random exogenous influences. In this model, the $\beta$ and $\theta$ parameters measure, respectively, the marginal effect of average past increments and the marginal effect of average past shocks on the current increment. The larger these parameters, the more autocorrelated the food security trajectory. When fitting the data to estimate the parameters $(g,\beta,\theta,\sigma)$ of this model, parsimonious orders $(p,q)$ for the ARMA process can be identified using a standard Bayesian criterion such as the Akaike information criterion (AIC). In particular, low orders will be favored when the number of data points is small.\footnote{A rule of thumb requires a minimum of 50 points for ARMA modeling of empirical data to be meaningful \cite{Chatfield:2016uv}, though 100 points would be recommended by many analysts.} 

Within the framework of \eqref{ARMA}, food security trajectories can be analyzed in terms of four fundamental dynamical properties:
\begin{enumerate}
\item{The \textit{level} of the food security variable $k_t$ itself. We choose to measure level in the subsequent analysis by the mean of $k_t$ over the entire time series for each actor.}
\item{The \textit{trend} $g$ of the variable, i.e., the mean increment $\mathbb{E}(\Delta_t)$. We consider an actor with a positive (negative) trend as \textit{improving} (\textit{deteriorating}); when the trend is not significantly different from zero, we say that the time series is \textit{neutral}.} 
\item{The \textit{volatility} of the increment time series $\Delta_t$. We distinguish between the absolute volatility, the standard deviation $\sigma$ and the relative volatility  
\begin{equation}
\rho=\frac{\sigma}{2\vert g\vert}-1.
\end{equation}
 The $\rho$ metric compares the value of $\sigma$ to the long-term trend $g$. We consider that a time series is not volatile in relative terms when shocks are too small to invert the trend, i.e., when $\rho<0$. This condition requires time series with faster growth trends to face proportionally larger shocks in order to be considered volatile. Note that volatility contains inseparable information on both intrinsic shock magnitude and actor response to shocks.}
\item{The \textit{persistence} of shocks as captured by the increments $\Delta_t$, relative to their average value $g$. As noted earlier, persistence quantifies the extent to which present increments are partially determined by past increments, i.e., the extent to which the increments retain a memory of their lagged values. Complete information about such serial associations is provided by the autocorrelation function $\gamma(s)$, with $\gamma(s)$ the  correlation coefficient between the lagged increment $\Delta_{t-s}$ and the present increment $\Delta_t$. In many cases, however, a more compact measure of persistence is desirable. Within the ARMA framework of \eqref{ARMA},  $\beta_i$ represents the partial correlation coefficient between the present excess increment $(\Delta_t-g)$ and its lagged value $(\Delta_{t-i}-g)$, while $\theta_j$ is the partial correlation coefficient between  $(\Delta_t-g)$ and past shocks $\epsilon_{t-j}$. This suggests the following simple definition for overall persistence:\footnote{Though among the simplest definitions available, the particular definition of persistence given in \eqref{persistence} is not the only one possible. Alternatives in the literature include the power spectrum at zero frequency, variance ratios \cite{Cochrane:1988gt}, mean reversion \cite{Dias:2010kc}, half-life, approximate entropy \cite{Pincus:1991ju}, and, for series with long-range autocorrelation, Hurst exponents and detrended fluctuation analysis \cite{Peng:1994be}.} 
\begin{equation}\label{persistence}
	\pi=\sum_{i=1}^p\beta_i+\sum_{j=1}^q\theta_j. 
\end{equation}
We call an actor's response to shocks \textit{anti-persistent} if $\pi<0$, \textit{random} if $\pi=0$, and \textit{persistent} if $\pi>0$.} As noted before, the condition that $\pi<0$ can be understood intuitively as expressing that the fluctuations of food security increments $\Delta_t$ are quickly reversed---shocks do not last. A time series exhibiting this property is more predictable in the long run than one for which shocks have persistent effects ($\pi>0$). Observe, however, that high persistence is not necessarily associated with high volatility: the two quantities are independent features of a time series. See also Figs. \ref{resilience}-\ref{resistance}.	
\end{enumerate} 

Several remarks are in order. First, structural breaks in the trend (e.g., caused by wars, massive aid flows, and other protracted events) should be identified \textit{a priori} and taken into account when fitting the welfare variation trajectory with an ARMA model. In particular, persistence can be overestimated when such breaks are present in the data \cite{Perron:2006wv}; an actor with a trend that is best approximated by a spline or other nonlinear function will be misinterpreted by the ARMA model in \eqref{ARMA} as one with a linear trend but large, long-lasting deviations away from this trend. Second, in most applications the order $(p,q)$ will be small ($p+q\leq 2$). We highlight two special cases: when $p=q=0$, the food security trajectory cannot be distinguished from a random walk with drift (a ``stochastic trend'' or ``unit root''); when $(p,q)=(0,1)$ and $\theta_1=-1$, the trajectory consists of a linear trend with independent shocks (a ``deterministic trend''). Third, the constant $1/2$ in the definition of $\rho$ is to some extent arbitrary: its value is contingent upon a quantification of $\textrm{Prob}(g+\epsilon_t<-g)\ll 1$. Here we use the common ``one sigma'' definition. Fourth, while ARMA modeling provides a solid, well-understood framework for understanding serial correlations, more general formulations (e.g., autoregressive fractionally integrated moving average (ARFIMA) models or autoregressive conditionally heteroskedastic (ARCH) models, non-constant trends) can be used if required, provided the number of data points is large enough to allow for such refinements. Fifth, trend, volatility, and persistence can also be defined non-parametrically, e.g., from the empirical mean, standard deviation, and autocorrelation coefficients of the increment series $\Delta_t$. These alternative definitions can be used as consistency checks.\footnote{Note that the shocks $\epsilon_t$ need not be normally distributed in general. Rather, the skewness $S$ and kurtosis $K$ of the fit residuals should be thought of as providing additional information about the underlying welfare dynamics. An actor with positive residual skewness (such as Montenegro, $S=2.75$) and another with negative residual skewness (such as Croatia, $S=-5.02$)  for instance, have qualitatively different  trajectories. The former undergoes more frequent negative shocks; when positive shocks to growth occur, however, they tend to have a larger magnitude. Analyzing departures from normality can thus be part of a finer-grained analysis of welfare trajectories.} Finally, note that shocks themselves do not need to be observed in order to quantify resilience; time series data on food security outcomes is sufficient. Given the difficulty of measuring the inherent magnitude of shocks, this is an important feature of our approach.

	
With these observations in mind, we propose the following formal definition of food security resilience: \textit{an actor is resilient if the long-term food security trend is not deteriorating ($g\geq 0$) and food security increments are negatively correlated in time, i.e., exhibit anti-persistence} ($\pi<0$; see Fig. \ref{resilience}). Secondarily, we can think of an actor as \textit{resistant} to shocks if the long-term food security trend is not deteriorating  ($g\geq 0$) and the fluctuations in increments do not indicate volatility ($\rho<0$; see Fig. \ref{resistance}.) These definitions mathematically express our suggestion that a desirable food security trajectory is one which can resist and recover from shocks along a generally non-negative trend.

\section{An analysis of country kilocalorie availability}

We now illustrate these concepts with a real-world dataset: annual per capita kilocalorie (kcal) availability between 1961 and 2011 for 161 countries in the world, taken from the FAOSTAT database \cite{FAOSTAT:ufZUrcWn}. Note that our approach can accommodate any quantitative variable at any scale. We choose national-level kcal availability because this indicator is one of the most common proxies for food security. However, other welfare variables for which high-frequency data is available, including those at household level, could also be used.

The kcal dataset is constructed using food balance sheets that add domestic agricultural/livestock production to imports, and then subtract exports, livestock feed, seed, and losses during storage and transport. The annual net totals are converted to kilocalories and then divided by the country's population in each year to obtain per capita kcal availability. Note that this figure does not capture the often pronounced distributional inequalities within countries. In addition, kcal consumption is only one aspect of food security, which also encompasses nutrient intake, food safety, cultural preferences, and other dimensions  \cite{Vaitla:2015wn}. At best, per capita kcal availability figures can be thought of a coarse upper bound estimate of a single aspect of country-level food security. 

We first observe that kcal availability $k_t$ is far from being stationary (Fig. \ref{global_evolution}). Most countries increased their kcal availability level in the last five decades, although fifteen countries had lower levels in 2011 than they did in 1961. These declines occurred despite robust growth in the global average over this period; the entire series has a strong upward trend, broken only temporarily by a five-year interval corresponding to the breakup of the Soviet bloc in the late 1980s and early 1990s. An important feature of the global distribution of kcal availability between $1961$ and $2011$ is convergence: while the $1961$ kcal distribution was bimodal, the $2011$ distribution is unimodal (Fig. \ref{distributions}). However, the distributions of $k_t$ remain significantly positively skewed and platykurtic at all times, indicating that global inequalities in food availability remain high. Using the Dickey-Fuller F test, a unit root can be excluded at the $1\%$ level for only six countries (Belgium, Ecuador, India, Panama, Sweden, and Switzerland), indicating that $k_t$ series are usually not consistent with trend-stationary processes. This is in itself a notable finding: a shock to kcal growth usually has long-lasting impact on the levels of $k_t$---that is, $k_t$ does not merely revert to its original trajectory. No significant multi-year cycles were observed.

\begin{figure*}[t!]\centering
	\includegraphics[scale=.8]{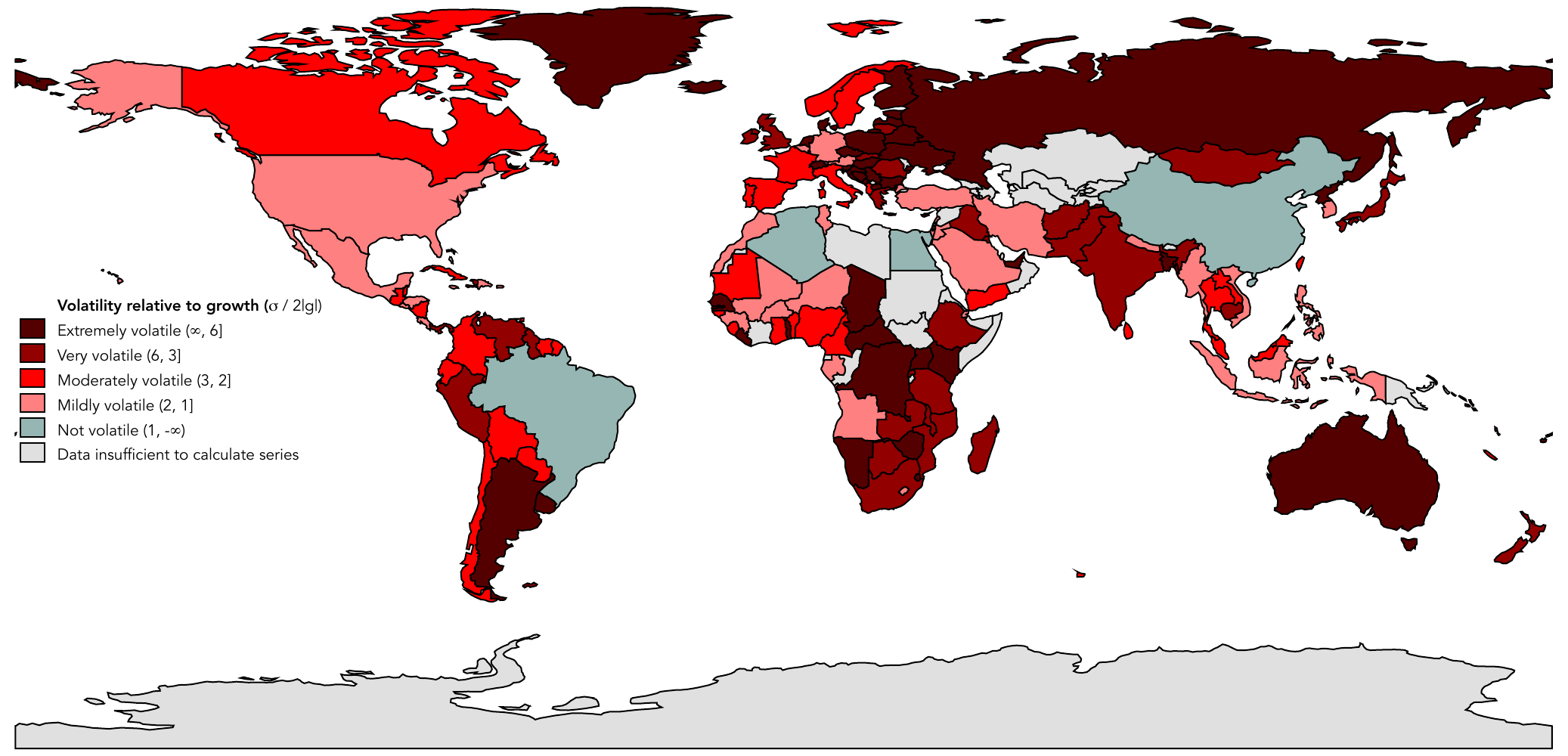}
	\caption{Volatility of per capita daily kilocalorie availability, based on annual data 1961-2011. Trend inversions $g\mapsto -g$ are unlikely if the mean fluctuation $\sigma$ is less than twice the size of the long-term growth trend ($\sigma<{2\vert g\vert}$). All four countries labeled in green---Egypt, China, Algeria, and Brazil---also have improving long-term trends, and so are considered resistant.}
	\label{rhomap}
\end{figure*}

\begin{figure*}[t!]\centering
	\includegraphics[scale=.8]{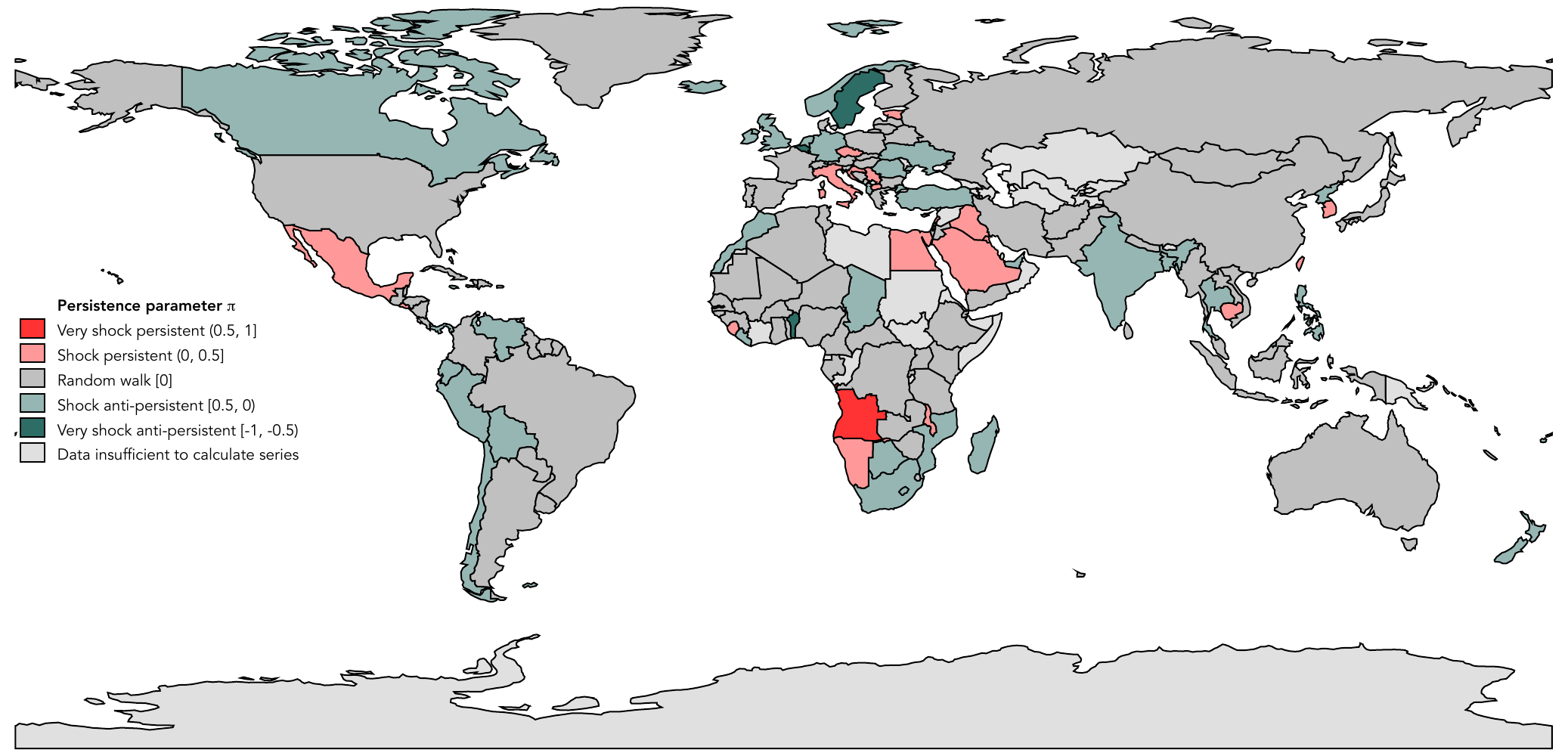}
	\caption{Persistence of per capita daily kilocalorie availability, based on annual data 1961-2011. $\pi>0$ (red) suggests persistence; $\pi=0$ (gray) suggests a random walk; $\pi<0$ (green) suggests anti-persistent behavior. Cambodia's and Macedonia's high persistence are not statistically significant (Table S1).  \textbf{All countries colored green except Chad and Madagascar, both of which have a declining long-term kilocalorie trend, are resilient.}}
	\label{pimap}
\end{figure*}

Next we performed the regression \eqref{ARMA} with AIC-selected orders $(p,q)$ after checking that the increment time series $\Delta_t$ in \eqref{ARMA} are stationary (in the Dickey-Fuller sense). Figs \ref{rhomap}-\ref{pimap} show the performance of all 161 countries in our dataset between 1961-2011 for the volatility parameter $\rho$ and the persistence parameter $\pi$; similar maps for mean kilocalorie levels and long-term kilocalorie trends are given in the SI (Fig. \ref{levelmap}-\ref{trendmap}). Fig. \ref{trendmap} shows marked geographical differences in the mean size of annual increments. With respect to developing world regions, the performances of West Africa, the Middle East, and Central America (as well as Brazil in South America) are notable, in addition to the well-known success stories of East and Southeast Asia. We see relatively worse performance in much of Central and Southern Africa, as well as South Asia.

The patterns of volatility and persistence are more unexpected. By the $\rho$ criterion---having an average fluctuation size that does not threaten inversion of the long-term trend---only four countries in the world are not volatile: Egypt, China, Algeria, and Brazil (\ref{rhomap}). This structural strength is mostly due to steeply positive long-term trends: Egypt, China, and Algeria have the highest $g$ in the sample, with an increment of around 30 kcal/year, and Brazil is not far behind at 22 kcal/year. These four countries are thus also resistant. However, most countries in the world, especially those with very flat trends (e.g., Russia, Argentina, countries in Eastern Europe) and/or high-magnitude fluctuations (Central and Southern Africa), frequently experience serious shocks.

In the presence of strong persistence of the effects of shocks, such high volatility can disrupt the long-term growth trajectory. We see, however, a slightly more positive global picture when looking at the persistence parameter $\pi$: only a handful of countries hold on to past trends. The trajectory of most of the world's countries, in fact, is best characterized as a random walk; their kilocalorie availability is neither dragged down by nor recovers from shocks. Note that this not an unambiguously positive characteristic---the experience of shocks in these countries delays return to pre-shock levels and trends, relative to anti-persistent countries.

No country in the entire sample is both resilient and resistant to shocks. However, 36 countries---all of those labeled in green in Fig. \ref{pimap}, with the exception of Chad and Madagascar---are resilient. Within this set are many of the least developed countries in the world, including (in order of decreasing $g$) Benin, Lesotho, Mozambique, Bangladesh, and Liberia. Given that many other countries at similar levels of development do not bounce back well from shocks, investigating the determinants of resilience is an important direction for future research.

\section{Implications for policy and research}

\begin{figure}[t!]\centering
	\includegraphics[scale=.9]{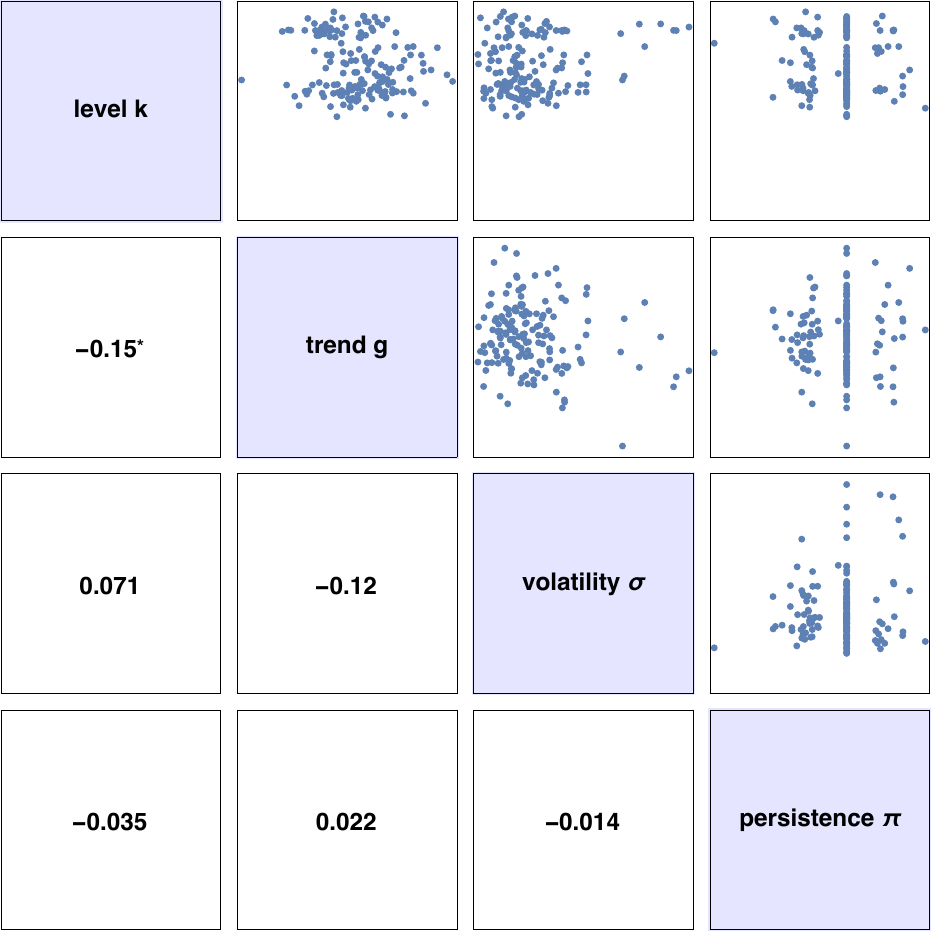}
	\caption{Independence of the four fundamental properties of food security trajectories for the country kilocalorie availability dataset. Corresponding insets are Spearman correlation coefficients. Except for a weak correlation between level and trend, all variables are uncorrelated at p<0.1.}
	\label{no_correlation}
\end{figure}

Three key policy and research implications emerge from this analysis. First, while previous works have examined the dimensionality of food security in terms of what different variables measure \cite{Barrett:2010fj, Maxwell:2014kd, Vaitla:2015wn}, we suggest that multiple interpretations of a \textit{single} dynamic variable are possible. Assessing food security change over time requires (at least) an investigation of 1) levels attained, 2) the trend of growth or decline, 3) the volatility of outcomes, and 4) persistence of the effects of shocks. From subsets of these, the qualities of resilience and resistance can be identified. Surprisingly, the four properties above are independent of each other, which points to the need to think of them as complements, not substitutes, in the diagnosis of food security trajectories (Fig. \ref{no_correlation}). This latter point is seen by comparing the ten best and worst performing countries in the kcal sample, with respect to each property (Table \ref{bestworstperf}).

Second, more research on the determinants of variation in persistence and volatility, and by extension resilience and resistance, is critical. We fit a simple linear model to test the hypothesis that trade openness is positively correlated to kilocalorie volatility and negatively correlated to persistence. Using mean per capita income, literacy rate, and democratization as controls, we find that trade openness is not significantly associated with either volatility or persistence.\footnote{We also fit a larger model including conflict deaths, oil revenue, and transport infrastructure, but none of these variables improved model performance, as evaluated either by overall goodness-of-fit or the magnitude and significance of individual parameters.} GDP per capita is negatively associated with persistence, but the magnitude of the effect is small. These same variables, however, do better in predicting mean kilocalorie levels (especially) and trends (Table \ref{regresults} in SI), again suggesting that distinct forces are driving the various properties of food security trajectories. More detailed work, including on the sub-country level, is needed; the approach outlined in this paper can be applied at any scale, including at the household level, where much of the key research on the determinants of food security is done.

Third, we note that a set of 18 countries are both volatile and exhibit persistence with respect to the effects of shocks. A subset of these---Angola, Cambodia, El Salvador, Iraq, Malawi, Mexico, Namibia, and Sierra Leone---have large segments of the population with low kilocalorie intake. Livelihoods and health are likely to be severely impacted by shocks in these areas, and recovery is likely to be protracted. From a resilience perspective, these are priority countries for international donors seeking to invest in bolstering the ability of the state and the market to prevent and mitigate the impact of shocks on kilocalorie intake. Global food security monitoring systems, especially those based on high-frequency long-term sentinel sites \cite{Headey:2015dg}, are critical to understand why these countries, but not others with similar levels or trends of food security, are particularly volatile and shock-persistent.

In conclusion, we note that, for policymaking purposes, a deeper investigation of persistence and volatility may improve forecasting of future trends in welfare. In a retrospective fitting of the ARMA model presented here, the trend, volatility, and persistence parameters move independently. More generally, however, the stronger the persistence of shock effects, the more difficult future forecasting of a trend will be; a shock occurring in a persistent country will be magnified, as deviations from the past trend will last longer.\footnote{High persistence may be desirable in the short-term during periods of rapid positive growth. However, the detrending procedure and construction of the ARMA model requires that persistence during periods of positive growth is mirrored by persistence during periods of decline over the extent of the time series examined. The implications of this trade-off for very long-term food security (i.e., longer than the time series in question) is unclear. Overall, high persistence generally indicates a structural weakness in the ability to be free from the impact of past fluctuations.}  In contrast, a country that is structurally anti-persistent will tend to revert back to the trend more predictably. Similarly, the more volatile the fluctuations, the more difficult it will be to predict the sign of the future trend. Data limitations may complicate country-level analysis of food security, but household- and individual-level time series can help illuminate these issues.

\acknowledgments{Research at the Perimeter Institute is supported in part by the Government of Canada through Industry Canada and by the Province of Ontario through the Ministry of Research and Innovation. Thanks to Chris Golden, Robert Gustafson, E. Toby Kiers, Janet Kim, Erwin Knippenberg, William Masters, Dan Maxwell, Mark Constas, Beth Pringle, Ben Rice, Elizabeth Stites, Patrick Webb, and Ahmed Youssef for valuable feedback on the manuscript.}


\newpage
\bibliographystyle{alpha}
\bibliography{resilience,library}

\clearpage

\begin{center}
	\huge{Supporting Information}
\end{center}

\appendix

\section{Per capita kilocalorie database}

See file kcal.csv for per capita kilocalorie availability database. Values for Belgium 1961-1999 and Luxembourg 1961-1999 are taken from combined Belgium-Luxembourg data. Values for Czech Republic and Slovakia 1961-1990 are taken from combined Czechoslovakia data. Values for Serbia and Montenegro 1991-2005 are taken from combined Serbia-Montenegro data. Values for Bosnia-Herzegovina, Croatia, Macedonia, Montenegro, Serbia, and Slovenia 1961-1990 are taken from Yugoslav SFR data. Values for Belarus, Estonia, Latvia, Lithuania, Russia, and the Ukraine 1961-1990 are taken from USSR data.

\section{Statistical properties database}

See file results.csv for statistical properties of country time series, including candidate ARMA models and the corresponding AIC scores. 

\clearpage

\section{Global distribution of kcal availability}

\begin{figure}[h!]\centering
	\includegraphics[width=.47\linewidth]{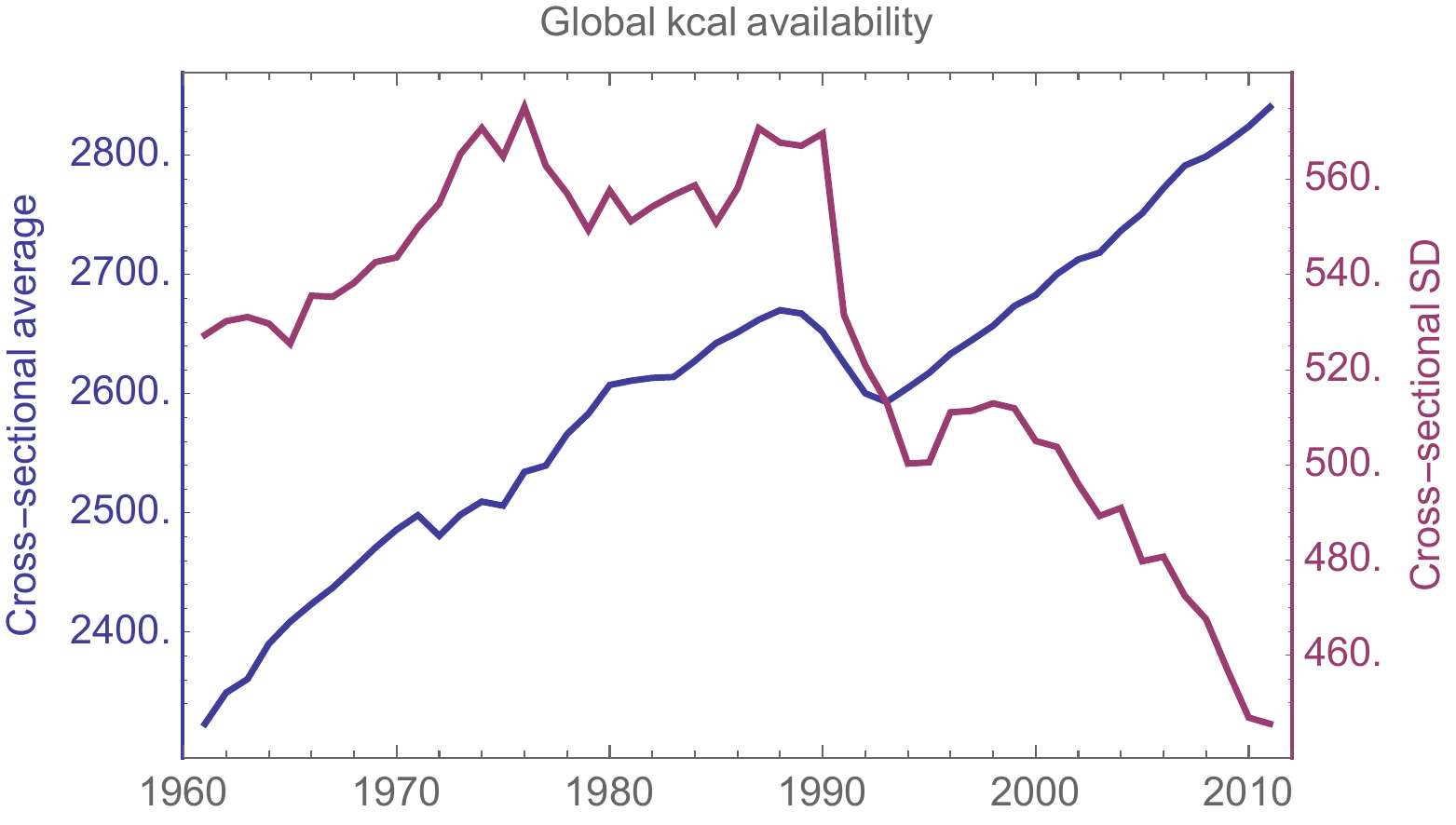}\hfill\includegraphics[width=.47\linewidth]{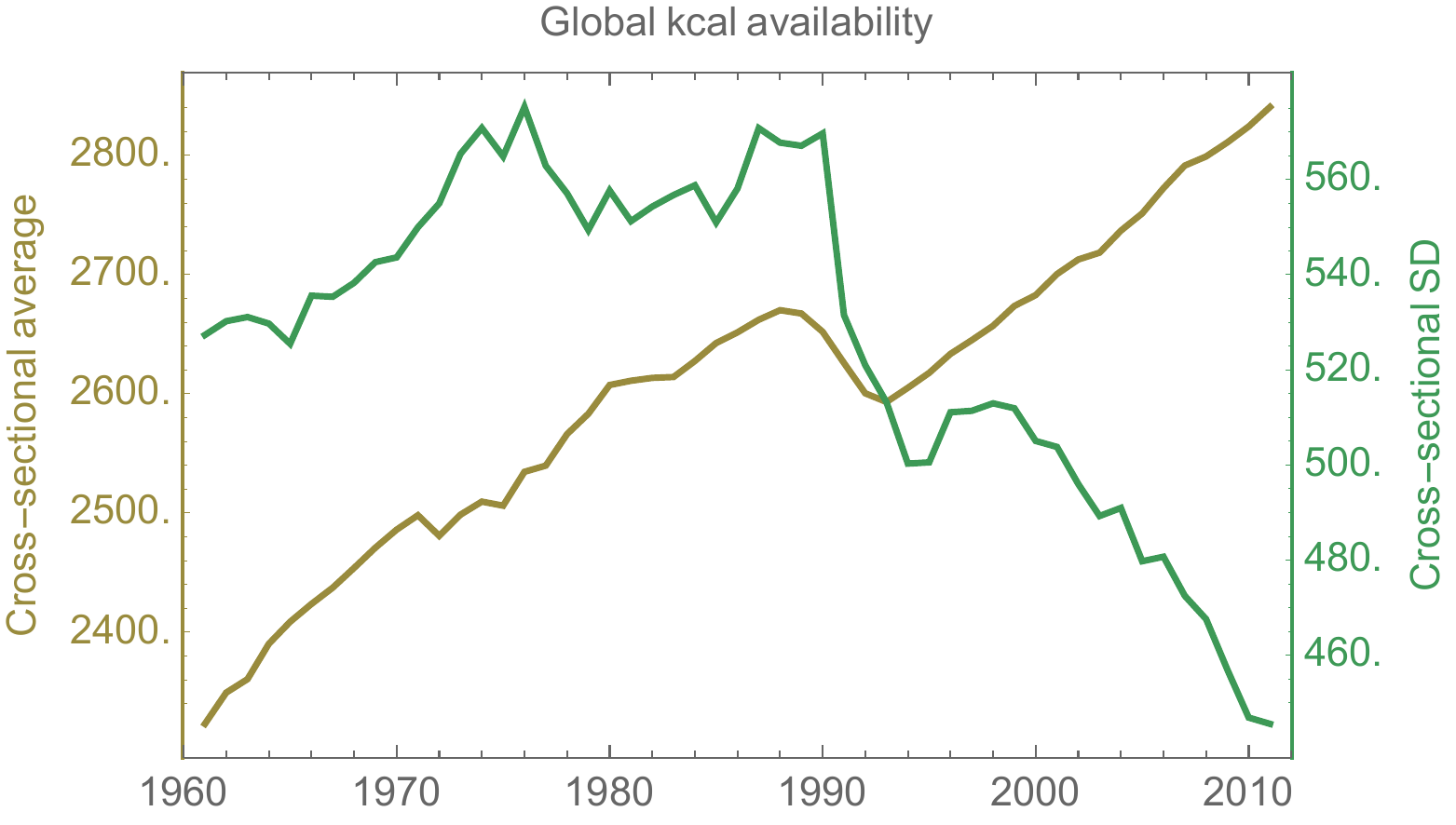}
	\caption{Left: Evolution of the global average (blue) and standard deviation (red) in country-level kcal availability for 1961 - 2011, indicating both growth and convergence during this period (with an exception around 1990). Right: Evolution of the sample skewness (blue) and sample kurtosis (red) for the same dataset, indicating non-normality of the global distribution of kcal availability.}
	\label{global_evolution}
\end{figure}

\begin{figure*}[h!]\centering
	\includegraphics[scale=.6]{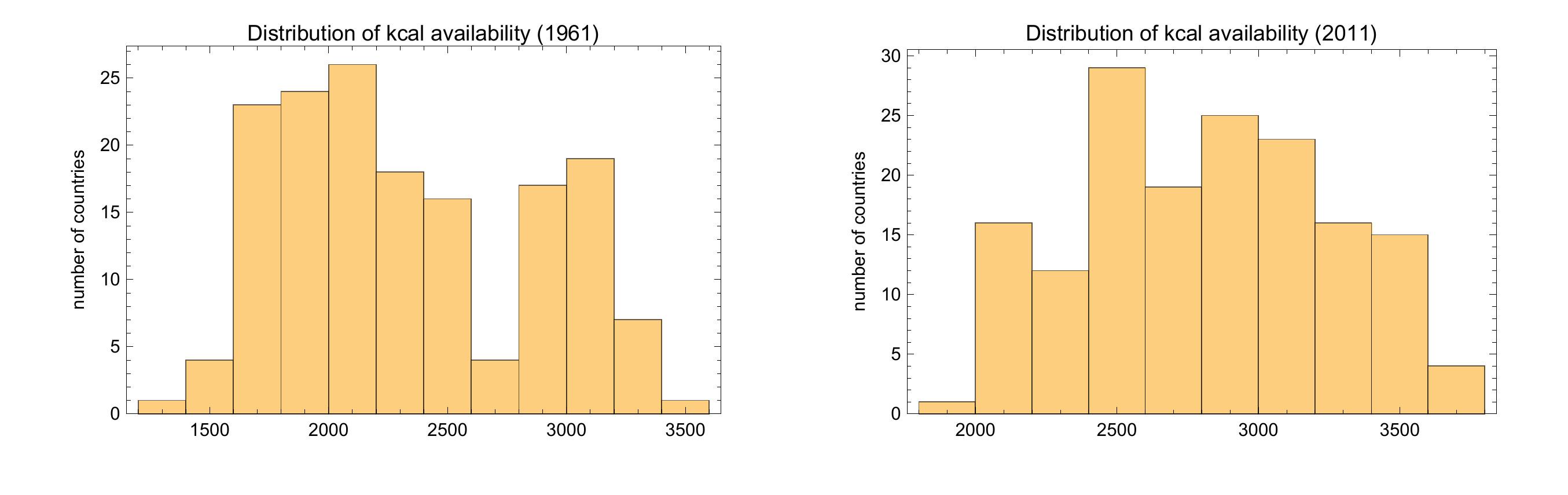}
	\caption{Convergence in food security: the distribution of kcal availability was bimodal in 1961 but became unimodal by 2011. This is consistent with a decrease in the cross-sectional standard deviation, see Fig. \ref{global_evolution} above.}
	\label{distributions}
\end{figure*}

%
\FloatBarrier

\newpage

\clearpage
\section{kcal availability levels and trends}

\begin{figure*}[!ht]\centering
	\includegraphics[scale=.75]{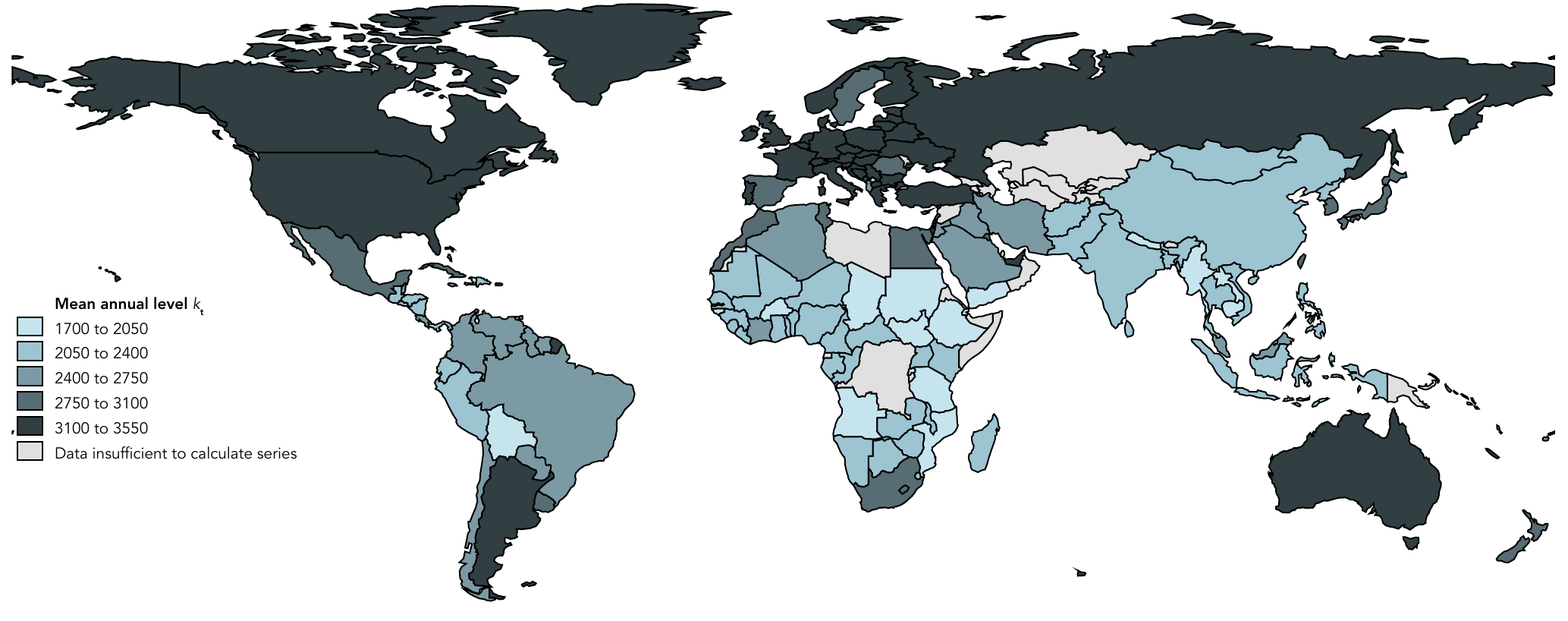}
	\caption{Levels of per capita daily kilocalorie availability. Mean of all annual values for each country between 1961-2011 is shown.}
	\label{levelmap}
\end{figure*}

\begin{figure*}[!ht]\centering
	\includegraphics[scale=.75]{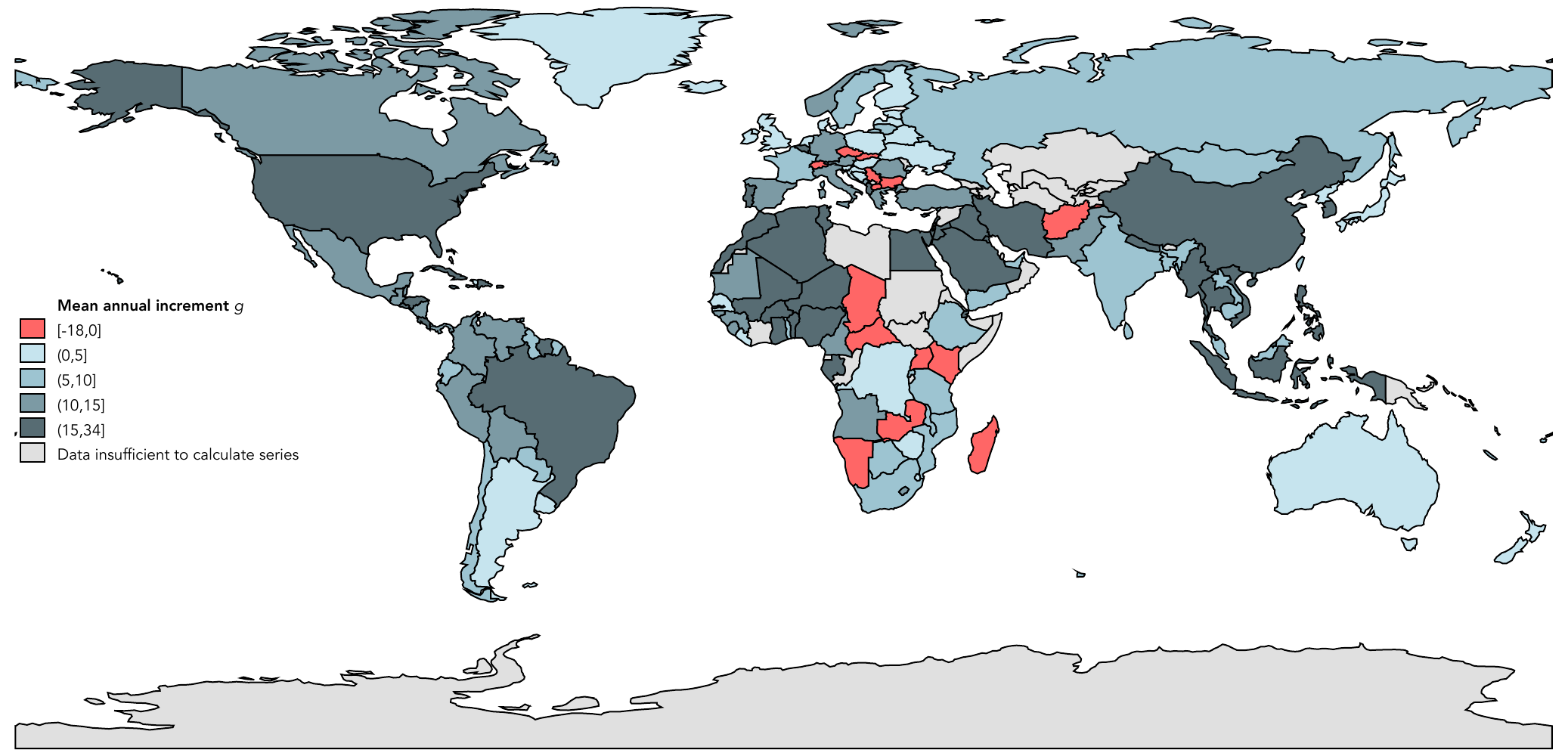}
	\caption{Trends of per capita daily kilocalorie availability. Mean of all annual increments in each country between 1961-2011 is shown.}
	\label{trendmap}
\end{figure*}

\FloatBarrier

\clearpage
\section{Countries ranking}

\begin{table*}[h!]
\begin{center}
	\begin{tabular} {| c || c | c | c | c |} \hline
& \textbf{Decreasing Level} & \textbf{Decreasing Trend} & \textbf{Increasing Rel. Volatility} & \textbf{Increasing Persistence} \\ 
\hline
1 & Turkey & China & Egypt & Sweden* \\
2 & Montenegro & Algeria & China & Benin*** \\
3 & Israel & Egypt & Algeria & Belgium*** \\
4 & Bosnia-Herzegovina & Saudi Arabia & Brazil & Luxembourg*** \\
5 & Belarus & Burkina Faso & Costa Rica & Chile*** \\
6 & Serbia & Iran & El Salvador & Thailand** \\
7 & Macedonia & Morocco & South Korea & Fiji* \\
8 & Ukraine & Mali & Iran & Ecuador*** \\
9 & Croatia & South Korea & Morocco & New Zealand*\\ 
10 & Russia & Cuba & Honduras & United Kingdom* \\
\vdots & \vdots & \vdots &\vdots &\vdots \\
152 & Cambodia & Central African Republic & Swaziland & South Korea*** \\
153 & Burkina Faso & Namibia & Slovenia & Macedonia\\
154 & East Timor & Macedonia & Macedonia & Estonia** \\
155 & Chad & Chad & Argentina & Serbia* \\
156 & Haiti & Zambia & Ukraine & Cambodia \\
157 & Mozambique & Bulgaria & Uganda & Kuwait*** \\
158 & Angola & Serbia & Bosnia-Herzegovina & Iraq** \\
159 & Djibouti & Madagascar & Finland & Sierra Leone** \\
160 & Myanmar & Slovakia & Czech Republic & Saudi Arabia***\\ 
161 & Ethiopia & Afghanistan & Croatia & Angola***\\
\hline
\end{tabular}
\end{center}
\caption{Best and worst performers by statistical feature, per capita daily kcal availability, 1961-2011. Stars in the last column indicate the $p$-value for autocorrelation of $\Delta_t$ according to the Ljung-Box test (***Significant at $p<0.01$; **significant at $p<0.05$; *significant at $p<0.1$).}
\label{bestworstperf}
\end{table*}

\FloatBarrier
\clearpage

\section{Cross-country regressions}

We fit the simple linear model 
\begin{equation}
	\rho, \pi = a + b*(trade) + \sum_{k=1}^{m} c_{i}X_{i} + \epsilon_{i}
\end{equation}
to test the hypothesis that trade openness ($trade$) is positively correlated to volatility relative to trend ($\rho$) and negatively correlated to persistence ($\pi$), given a vector of control variables $X_{1}....X_{m}$ representing economic resources, human capital stocks, and political participation (see Table \ref{regresults}). We also include the same determinants in models predicting mean kcal level and growth rate $g$. In all models, errors are independent and identically distributed with mean zero and standard deviation $\sigma$; error variances are heteroskedastic, and Huber-White standard errors are used in estimation. We also specify the same models using the subset of developing countries (as classified by the United Nations in 2012) only (Table \ref{regresults2}); results are similar. 

\bigskip

\begin{table} [!htbp]
	\centering
		\resizebox{\linewidth}{!}{
	\begin{tabular} {| l || l | l | l | l |} 
	\hline
	& $kcal$ (level) & $g$ (trend) & $\rho$ (volatility relative to trend) & $\pi$ (persistence) \\ \hline
	trade & -0.382 (0.908) & -0.046** (.018) & -0.083 (0.068) & -0.000 (0.001) \\ 
	gdppc & 0.010*** (0.002) & -0.000 (.000) & 0.000 (0.000) & -0.000* (0.000) \\ 
	literacy & 1006.058*** (185.671) & 4.862 (6.395) & -64.301 (59.835) & 0.145 (0.112) \\ 
	polity & 16.185** (6.206) & -0.509*** (0.161) & -0.248 (0.617) & -0.009* (0.004) \\ \hline
	constant & 1596.485 (128.69) & 9.360 (5.204) & 37.440 (36.119) & -0.124 (0.088) \\
	$r^{2}$ & 0.653 & 0.107 & 0.010 & 0.084 \\
	\hline
	\end{tabular}}
	\caption{Results of all country models predicting level, trend, volatility relative to trend, and persistence of per capita daily kcal availability. Standard errors in parentheses. ***Significant at p<0.01; **significant at p<0.05; *significant at p<0.1. Sources: trade openness (sum of exports and imports of goods and services measured as a share of gross domestic product) \cite{WB:2016wd}; GDP per capita, literacy \cite{10.1257/aer.20130954}; polity10 (degree of democracy and autocracy) \cite{Marshall:2013vk}.}
	\label{regresults}
	\end{table}

\begin{table} [!htbp]
	\centering
		\resizebox{\linewidth}{!}{
	\begin{tabular} {| l || l | l | l | l |} 
	\hline
	& $kcal$ (level) & $g$ (trend) & $\rho$ (relative volatility) & $\pi$ (persistence) \\ \hline
	trade & -0.072 (1.151) & -0.039** (.022) & 0.039 (0.089) & -0.000 (0.001) \\ 
	gdppc & 0.017*** (0.003) & -0.000 (.000) & -0.000 (0.000) & -0.000 (0.000) \\ 
	literacy & 700.963*** (177.437) & 8.193 (6.852) & -62.530 (59.597) & 0.071 (0.119) \\ 
	polity & 15.620** (7.379) & -0.500*** (0.183) & -0.881(1.093) & -0.010** (0.004) \\ \hline
	constant & 1785.826 (124.801) & 7.158 (5.442) & 30.299 (33.334) & -0.105 (0.089) \\
	$r^{2}$ & 0.512 & 0.085 & 0.018 & 0.064 \\
	\hline
	\end{tabular}}
	\caption{Results of developing country only models predicting level, trend, relative volatility, and persistence of per capita daily kcal availability. Standard errors in parentheses. ***Significant at $p<0.01$; **significant at $p<0.05$; *significant at $p<0.1$. Same sources as Table \ref{regresults}.}
	\label{regresults2}
	\end{table}
	

\FloatBarrier
\clearpage

\section{Illustration of resilience and resistance with synthetic data}

\begin{figure*}[h!]\centering
	\includegraphics[scale=.45]{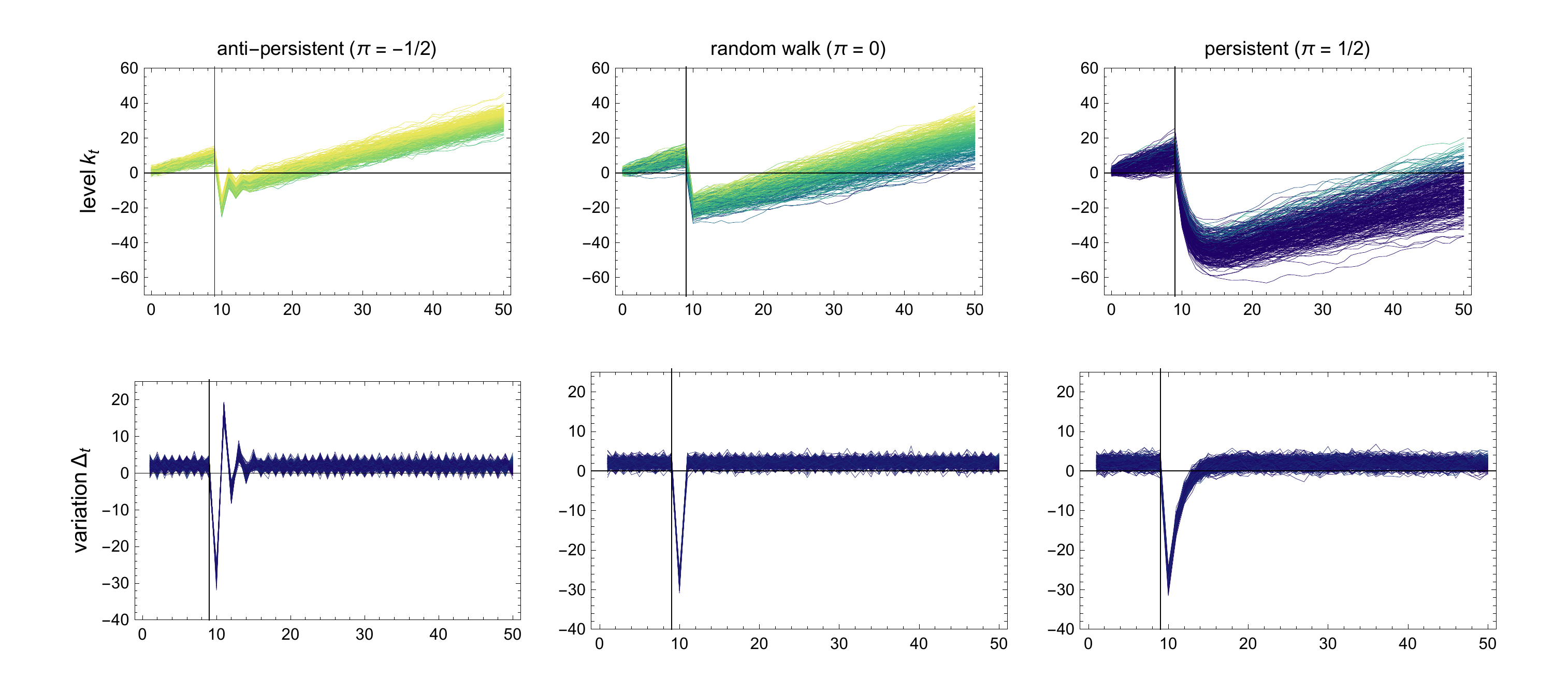}
	\caption{Resilient and non-resilient trajectories, constructed with synthetic data. The actor in the left column, with $\pi<0$ and a non-deteriorating long-term trend, is resilient. The actor in the right column, with $\pi>0$, exhibits persistent, and thus non-resilient, behavior. The actor in the middle column corresponds to the marginal case of random walk with drift.}
	\label{resilience}
\end{figure*}

\begin{figure}[h!]\centering
	\includegraphics[scale=.45]{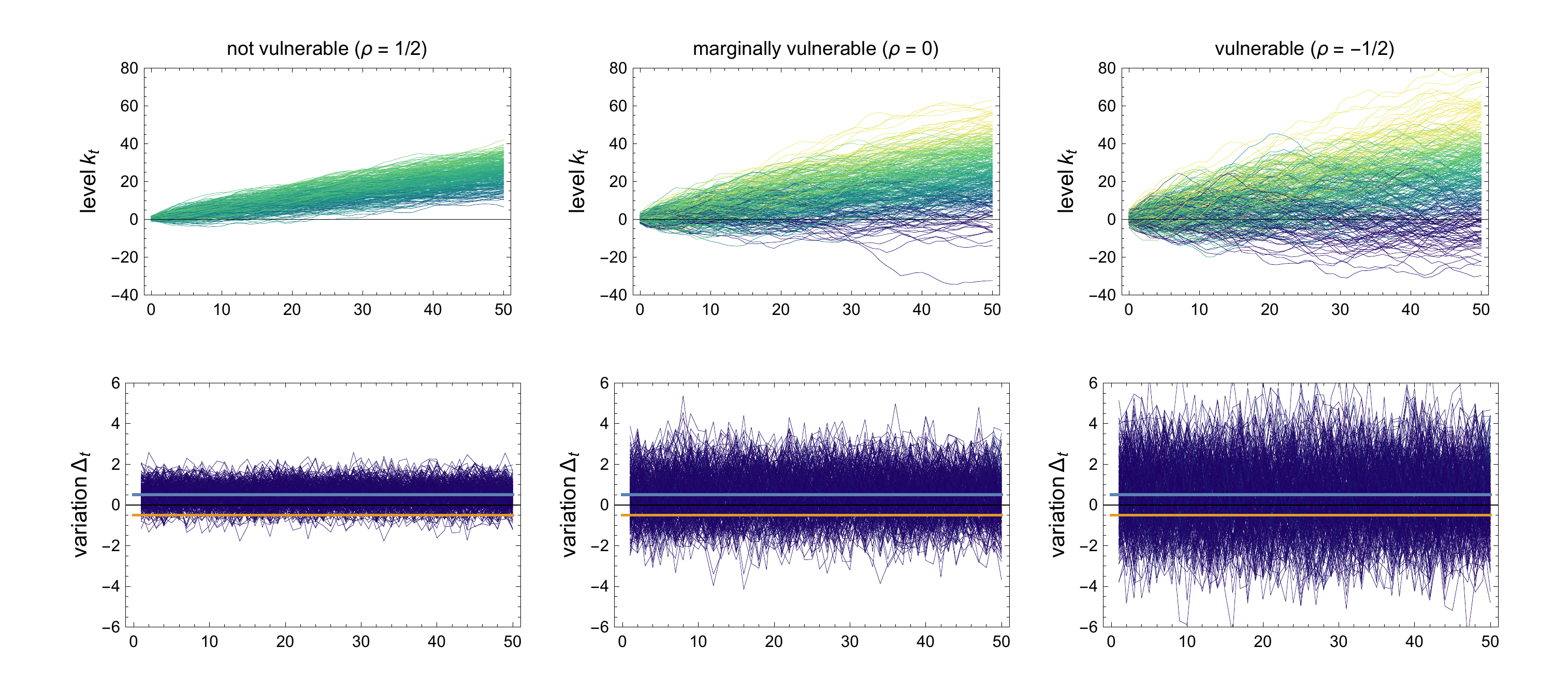}
	\caption{Resistant vs. volatile trajectories. The actor in the left column, with $\rho\geq0$ and a non-deteriorating long-term trend, is resistant to shocks. The second and third rows illustrate that volatile actors experience much stronger fluctuations. Here the blue line indicates the mean trend $g$, while the orange line gives the opposite trend $-g$.}
	\label{resistance}
\end{figure}

Below we illustrate our concepts of resilience and resistance  using synthetic data ($300$ realizations) generated with different ARIMA(1,1,0) model with identical trends but varying autoregressive coefficient $\beta_1$ (Fig. \ref{resilience}) or absolute volatility $\sigma$ (Fig. \ref{resistance}). 

\end{document}